\documentstyle[aps,prl,epsf]{revtex}

\begin{document}

\twocolumn

\title{Freezing of polydisperse hard spheres}

\author
{ N. G. Almarza and E. Enciso}

\address{
Departamento de Qu\'{\i}mica F\'{\i}sica I,
Facultad de Ciencias Qu\'{\i}micas
Universidad Complutense, \\
E-28040 Madrid, Spain}

\date{\today}
\maketitle

\begin{abstract}
The fluid - crystal equilibria of polydisperse mixtures of
hard spheres have been studied by computer simulation of the
solid phase and using an accurate equation of state for the fluid.
A new scheme has been developed to evaluate the composition
of crystalline phases in equilibrium with a given polydisperse fluid. 
Some common assumptions in
theoretical approaches and their results are discussed
on the light of the simulation results. Finally,
no evidence of the existence of a 
terminal polydispersity in the fluid phase is found for
polydisperse hard spheres, the disagreement of this finding 
with previous molecular simulation results is explained
in terms of the inherent limitations of some ways of
modeling the chemical potential as a function of
the particle size.
\end{abstract}

\vspace{1cm}
PACS numbers:
05.70.Fh,
64.70.Do,
82.70.Dd

The phase behavior of polydisperse mixtures of hard spheres (PHS)
has received some attention in recent years. 
Different theoretical approaches \cite{sear,barlet1,barlet2} 
and simulation methods \cite{bolhuis,bolhuis2,pmd}
have been used to gain knowledge about the transition from a 
polydisperse fluid phase to crystal phase(s). 
The theoretical approaches \cite{sear,barlet1,barlet2} use
to involve drastic approximations regarding the composition of the
phases, and the results are often presented in form of
stability diagrams \cite{sear,barlet1}, both facts should be 
carefully taken into account when interpreting the results.
Bolhuis and Kofke \cite{bolhuis} have studied the fluid - solid equilibria
by using molecular simulation methods, 
finding a "terminal polydispersity" in the fluid phase
that they interpreted as the maximum polydispersity of a fluid
which can originate a freezing transition, and related such a result
with some experimental data.
The origin of the "terminal polydispersity" in  Ref
\cite{bolhuis,bolhuis2} will be addressed in this work.

Let $P(\sigma)$ be a given
probability distribution function of particle diameters (PDFD). 
The distribution can be characterized
by its moments, $m_k = < \sigma^k >/\sigma_0^k $,
where $\sigma_0$ is a reference diameter.

The thermodynamics of PHS fluids is very accurately described by the
generalization of Salacuse and Stell \cite{salacuse} to the polydisperse
case of the equation of state (EOS)
due to Boublik and Mansoori, Carnahan, Starling and Leland (BMCSL)
\cite{bou,mcsl}.
In such an equation the pressure, $p$,  can be written as:
$\beta p =  \beta p ( m_1, m_2, m_3, \eta )$ where $\eta$ is the
packing fraction: $\eta = \pi N m_3 \sigma_0^3 /(6V)$. 
$N$ is the number of particles,
$V$ is the volume.
and $\beta \equiv 1/(k_B T )$, 
with $k_B$ being the Boltzmann's constant
and T the absolute temperature.

The excess chemical potential, $\mu_{ex}$, 
in the fluid phase takes the form:
$\beta \mu_{ex} \left( \sigma   \right)  =
\sum_{k=0}^3 c_k \sigma^k$, 
where the coefficients $c_k$ depend
on $m_1$, $m_2$, $m_3$ and either $\eta$ or $\beta p \sigma_0^3$.

The goal of the present work is to evaluate the fluid - solid equilibrium
for a given PDFD in the fluid phase. This point of view is
the main difference with the calculations of ref \cite{bolhuis},
however the statistical mechanics
underlying both procedures is basically the same.

In order to study polydisperse systems is convenient to
make use of the semigrand (SG) ensemble\cite{bolhuis,sg1}, where
the pressure, the total number of particles, $N$, and 
the chemical potential differences between 
the different species and a reference
one are fixed. For hard body interactions the
basic thermodynamic differential relation reads,

\begin{equation}
d \left[ N \beta \mu_0 \right]
= V d \left( \beta p \right)
+ \beta \mu_0 d N 
- \sum_{i \ne 0 } N_i d \left( \beta \mu_{i0} \right)
\end{equation}
where 
$\mu_0$ is the chemical potential of the reference species.
The sum is done over the other components, 
$N_i$ is the number
of particles of species $i$ and $\mu_{i0} \equiv \mu_i - \mu_0$.
Later,  a continuous distribution of sizes will be used 
however the discrete description is kept, 
for the shake of clarity in the equations:

An imposed chemical potential distribution (ICPD) is used to perform
the calculations, such a distribution should produce the
required PDFD  in the fluid phase.
As a difference with
the procedure in Ref \cite{bolhuis} here the ICPD
will depend on the pressure.
We have taken advantage
of the accuracy of BMCSL EOS. Such an equation let us
to link the fluid phase composition with the chemical
potential distribution at a given pressure.

Let $P_0(\sigma)$ be the expected PDFD, for instance, a
Gaussian distribution centered at $\sigma_0$ and with
standard deviation $\sigma_0 \lambda$. We can use
as input the values of $\beta \mu \left( \sigma \right)$
given by:

\begin{equation}
\label{bmu}
\beta \mu \left( \sigma \right) = 
- \frac{\left ( \sigma - \sigma_0 \right)^2 }{ 2 \sigma_0^2 \lambda^2 }
+ \beta \mu_{ex}^{BMCSL}
\left( \sigma, \beta p, \lambda \right)
\end{equation}
 
The actual PDFD of the fluid, $P(\sigma)$ will 
be practically identical
to $P_0$ due to the accuracy of the BMCSL EOS:
The two contributions to the chemical potential
can be grouped, by using an unique set of coefficients
$\{a_k \}$:
\begin{equation}
\beta \mu \left( \sigma \right) = 
\sum_{k=0}^3 a_k \left( \frac{\sigma}{\sigma_0} \right)^k
\end{equation}
The coefficients $a_k$ will be functions of $\lambda$ and $\beta p$.
The strategy to evaluate the fluid - solid 
equilibrium under the conditions stated above
lies on Gibbs-Duhem (or Clausius-Clapeyron) integration schemes
\cite{sg1,cc1,cc2}. The procedure is sketched
as follows: Starting for a given point ($\beta p_{0},\lambda_{0}$) in
which both phases are in equilibrium,
we obtain a trajectory on the $(\beta p, \lambda)$ plane that keep 
equilibrium conditions fulfilled. This can be done because
the chemical potential differences can be written as
functions of $\beta p$ and $\lambda$ through the coefficients $a_k$.
Therefore, considering $N$ fixed:

\begin{displaymath}
d \left[ N \beta \mu_0 \right] = 
\left[ V 
- \sum_{i \ne 0} N_i \left( 
\frac{ \partial  \left( \beta \mu_{i0} \right)}{\partial \beta p }
\right)_{\lambda} \right] d \left( \beta p \right)
+ 
\end{displaymath}
\begin{equation}
\left[- \sum_{i \ne 0} N_i \left( 
\frac{ \partial  \left( \beta \mu_{i0} \right)}{\partial \lambda }
\right)_{\beta p}  \right] d \lambda
\end{equation}
In the limit of a continuous 
distribution of sizes we can write
a Clausius Clapeyron analogue equation for the coexistence
$(\beta p, \lambda)$ line:

\begin{equation}
\left(\frac{ d \beta p}{d \lambda} \right)_{coex}
=
\frac
{\sum_{k=1}^3 
\left( 
\partial a_k / \partial \lambda 
\right)_{\beta p}
\Delta m_k }
{ \Delta  v  -   \sum_{k=0}^3 
\left( 
\partial a_k / \partial \left( \beta p \right) 
\right)_{\lambda}
\Delta m_k }
\end{equation}
where $\Delta$ represents the difference between the 
values of the corresponding property in the two phases and
$v \equiv V/N$.
The values of the derivatives of $a_k$ with respect to $\beta p$
and $\lambda$ can be evaluated numerically.

The starting point in the Clausius Clapeyron integration (CCI) was
the monodisperse hard sphere system 
($\lambda = 0$) where the equilibrium pressure
is known \cite{hoover,kranen} to be $\beta p \sigma^3 \simeq 11.71$. 
A second order predictor corrector
has been used to advance in the integration.
The integration step was  $\delta \lambda = 0.0025$,
the initial slope was found to be zero. 
 The fluid properties have been directly extracted from the
BMCSL EOS. A number of tests for several points on the
($\beta p, \lambda$) trajectory were performed by carrying out
SG Monte Carlo (SGMC) simulations on the fluid phase using $N=256$ 
and, within numerical accuracy, no differences between
simulation and theoretical results were found.
The solid phase was considered to be in a face centered cubic (FCC)
ordering and its properties were evaluated by SGMC simulation.

Details of the simulation procedure will be published 
elsewhere \cite{preparation}.
It suffices to say that three kind of moves were performed,
i) translation of the spheres (following the standard procedures),
ii) changes of a particle diameter,
by choosing the new diameter with probability proportional 
to $\exp[\beta \mu(\sigma)]$ with $\sigma \in [0, \sigma_{max} ]$,
$\sigma_{max}$ depends on the hard sphere interactions and
iii) Changes of volume, where
we found convenient to scale simultaneously the size of the
particles to enhance convergence on the sampling.

\begin{figure}[ht]
\begin{center}
\leavevmode
\epsffile{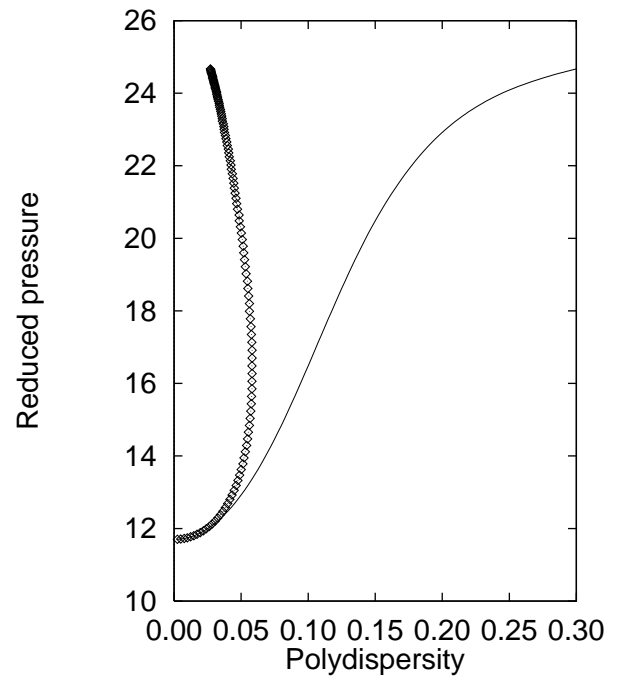}
\vspace{0.5cm}
\caption {Reduced pressure, $\beta p \sigma_0^3$ versus
the polydispersity of the two phases in equilibrium. 
The polydispersity,
$s_2$ is defined as: $s_2 = \left(m_2/m_1^2 - 1\right)^{1/2}$. For the
fluid phase $s_2 = \lambda$. Continuous line represents the result
for the fluid phase, symbols represent the conditions of
the crystal phase at equilibrium with the fluid as stated in the text}
\label{fig1}
\end{center}
\end{figure}

\begin{figure}[ht]
\begin{center}
\leavevmode
\epsffile{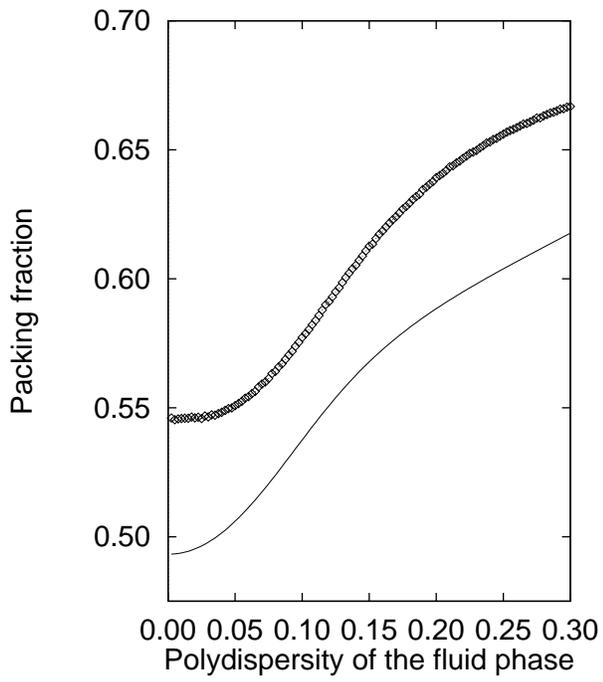}
\vspace{0.5cm}
\caption {Packing fraction of the fluid and crystal phases as
a function of $\lambda$ (see the text for details). Line
and symbols as in Fig. 1}
\label{fig2}
\end{center}
\end{figure}

\begin{figure}[ht]
\begin{center}
\leavevmode
\epsffile{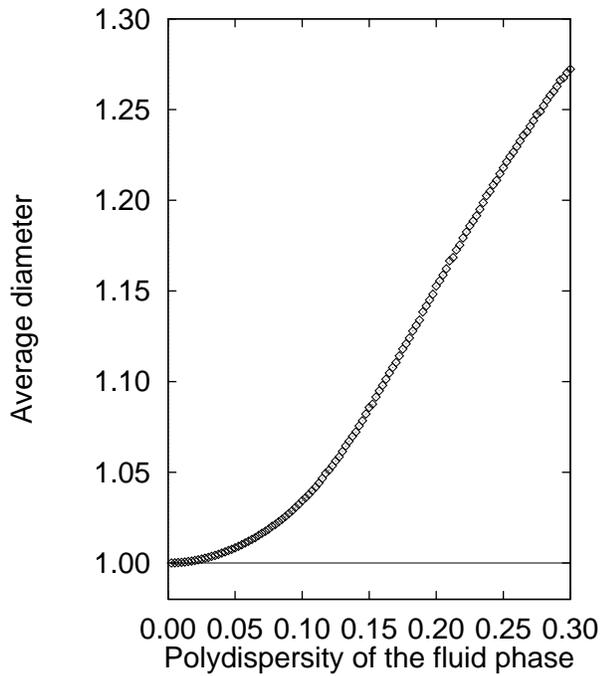}
\vspace{0.5cm}
\caption {Reduced average diameter of the samples, $m_1$ 
in different phases as a function of the fluid polydispersity
$\lambda$. Line and symbols as in Fig. 1}
\label{fig3}
\end{center}
\end{figure}

CCI were performed by systems with
with $N=108$ and $N=256$. No significant finite size
effects were found regarding the main conclusions
of the work. Some results for $N=256$ are presented
in figures.
In Fig \ref{fig1} the results of the pressure
as a function of the polydispersity in the two phases in equilibrium.
In figure \ref{fig2} we plot the packing fraction
of the coexisting phases as a function of the
polydispersity of the fluid phase.
In figure \ref{fig3} we show how the average
diameter of the crystal phase increases with the polydispersity
of the fluid phase.

In some theoretical work \cite{sear,barlet1} some conjectures have been
made regarding the possibility of finding a fluid
in equilibrium with two or more solid phases. This
result, based on diagrams of phase stability,
is not consistent with the phase rule, except in a number of
singular points.
The origin of such results lies in the theoretical approximation that
the fluid composition (polydispersity) is equal to the overall 
composition in the solid phases(s).
As can be seen in figures \ref{fig1} and \ref{fig3} such condition
is not fulfilled except for very low polydispersities.
However, one should not neglect the possibility of finding two
(or more) crystalline phases which could enter into
competition to become the solid phase in equilibrium
with the fluid. 
The phase diagram of the binary mixtures of hard spheres
\cite{kranen} can be used as an example;
In general for an equimolar fluid mixture $x=1/2$, the solid
phase in equilibrium with the fluid is richer in the large component.
As the size difference increases an eutectic point appears in the 
phase diagrams \cite{kranen},
it could happen that for given size difference
the eutectic composition could become $x_s=1/2$, in that case we 
could have for larger size differences,
that the stable solid phase could become
composed mainly by the small spheres.
In order to check such a possibility for the polydisperse system 
we have made some control simulations starting from FCC phases 
composed with particles smaller that $\sigma_0$ at pressures closed 
to the ones obtained in CCI.
Those systems evolved either to produce
the same solid appearing in the CCI
or to the melting of the sample.
These results seem to discard the change of stable phase in the procedure
of increasing $\lambda$ (at least in the range studied in this
work). It is clear, however, that such a competition between solid phases
could appear as the freezing proceeds, the change of the composition
of the fluid phase will alleviate the phase rule restrictions, and
it is quite likely to happen for high pressures where
the fluid could even disappear as an equilibrium phase.
Other possibilities have not been considered here, for instance,
the formation of crystal phases with a bimodal distribution of sizes,
which could be accommodated, for instance, in a 
body centered cubic lattice.

Here we will discuss briefly 
terminal polydispersity in the fluid phase
which appears in 
the results of Ref. \cite{bolhuis}. In that work
the ICPD has the form:

\begin{equation}
\label{icpdb}
\beta \mu_{i0} \left( \sigma \right) = 
- \frac{  \left( \sigma - \sigma_0 \right)^2}{ 2 \sigma_0 \lambda^2 }
\end{equation}
With this function a CCI scheme was performed
from the monodisperse limit ($\lambda = 0$), and it
was found that $m_1 \rightarrow 0$ when
$\lambda$ increases and that the values of the reduced polydispersity,
$s_2 = \sqrt{m_2/m_1^2 - 1}$, in the fluid at equilibrium
with a solid phase have to be less than a certain 
"terminal polydispersity", which the authors identified
with some experimental results of crystallization of colloidal mixtures.
In our simulations no such a terminal polydispersity appeared,
however the results are roughly similar to those of Ref 
\cite{bolhuis} until that point. 
This apparent anomaly can be explained in terms of
the form of ICPD. In our scheme, the excess contribution
to the chemical potential produces a positive value of the
coefficient $a_3$ of the ICPD, favoring
large diameters to compensate the effect of hard sphere
repulsions, however the form of the ICPD given by Eq (\ref{icpdb})
produces a limit in the maximum packing fraction of a fluid phase with
a certain polydispersity well below close packing.
It is possible to estimate such a limit by computer 
simulation \cite{bolhuis} or using the BMCSL EOS.
In fact the terminal points in the diagrams of \cite{bolhuis}
correspond just to the crossing between the fluid branch 
of the phase diagram and the line of such a maximum packing as a function
of the polydispersity. As pointed out in Ref \cite{bolhuis}
the end of the coexistence curve is conditioned by 
the ICPD, however such an end does not seem to correspond to any relevant
physical situation.

The existence of a terminal polydispersity in the fluid phase
has been also treated theoretically (See for instance
Ref. \cite{barlet2}), however the results are strongly
influenced by the restrictions to the size fractionation.

From the inspection of Fig. \ref{fig1} one could think that the stability
range of a polydisperse crystal with respect to the fluid
can lie in a range between two pressures. This is not the case,
we must emphasize that two points with the same value of the reduced
polydispersity on the crystal "branch" do not correspond to the same PDFD
(See Fig. \ref{fig2} and Fig.\ref{fig3}). 
The one corresponding to the higher pressure is associated with 
a distribution with greater value of $m_1$ and a more negative 
value of the skewness \cite{recipes} (i.e.
the distribution has an asymmetric
tail extending out towards small values of $\sigma$).

The authors acknowledge the financial support of DGICYT/Spain
(Grant No. PB95-072-C02-02)

\begin{thebibliography}{11}

\bibitem{sear}
R.P. Sear, Europhys. Lett. {\bf 44}, 531 (1998)

\bibitem{barlet1}
P. Barlett, J. Chem. Phys. {\bf 109}, 10970 (1998) (and references
therein)

\bibitem{barlet2}
P. Barlett and P.B. Warren, Phys. Rev. Lett. {\bf 82}, 1979 (1999)
(and references therein)

\bibitem{bolhuis}
P.G. Bolhuis and D.A.Kofke, Phys. Rev. E {\bf 54}, 634 (1996)

\bibitem{bolhuis2}
P.G. Bolhuis and D.A.Kofke, J. Phys.: Condens. Matter {\bf 8}, 
9627 (1996)


\bibitem{pmd}
S. Phan, W.B. Russel, J. Zhu and P.M. Chaikin, 
J. Chem. Phys. {\bf 108}, 9789 (1998=

\bibitem{salacuse}
J.J.Salacuse and G.Stell, J. Chem. Phys. {\bf 77}, 3714 (1982)

\bibitem{bou}
    T. Boublik,
     J. Chem. Phys., {\bf 53}, 471 (1970);


\bibitem{mcsl}
    G. A. Mansoori, N. F. Carnahan, K. E. Starling,  and
    T. W.  Leland,
    J. Chem. Phys., {\bf 54}, 1523 (1971).


\bibitem{sg1}
D.A. Kofke, in {\it Monte Carlo Methods in Chemical Physics},
Vol. 105 of {\it Advances in Chemical Physics},
edited by D.M.Ferguson, J.I.Siepmann and D.G.Truhlar, p. 405

\bibitem{cc1}
D.A. Kofke, {\em Mol. Phys.} {\bf 78}, 1331 (1993)

\bibitem{cc2}
D. Frenkel, in {\it Observation, Prediction and
Simulation of Phase Transitions in Complex Fluids},
Vol. 460 of {\it NATO Advanced Institute Series C:
Mathematical and Physical Sciences}, edited by M. Baus,
L.F. Rill and J.P.Ryckaert (Kluwer, Dordrechtm 1995), p.463

\bibitem{hoover}
W.G. Hoover and F.H.Ree, {\em J. Chem. Phys.} {\bf 49}, 3609 (1968)

\bibitem{kranen}
W.G.T. Kranendonk and D. Frenkel,
Mol. Phys. {\bf 72}, 679(1991)

\bibitem{preparation}
N.G. Almarza and E. Enciso (to be published)

\bibitem{recipes}
W.H. Press, B.P. Flannery, S.A. Teukolsky, and W.T. Vetterling,
{\it Numerical Recipes, The Art of Scientific Computing}
(Cambridge University, Cambridge, 1988)

\end{thebibliography}
\end{document}